# Topological Defects in the Random-Field XY Model and Randomly Pinned Vortex Lattices

Michel J. P. Gingras[1] and David A. Huse[2]

[1] *TRIUMF, 4004 Wesbrook Mall, Vancouver, B.C., V6T-2A3, Canada*

[2] *AT&T Bell Laboratories, Murray Hill, New Jersey 07974*

(February 6, 1995)

As a simplified model of randomly pinned vortex lattices or charge-density waves, we study the random-field XY model on square ($d = 2$) and simple cubic ($d = 3$) lattices. We argue, and confirm in simulations, that the spacing between topological defects (vortices) diverges more strongly than the Imry-Ma pinning length as the random field strength, $H$, is reduced. For $d = 3$ the data are consistent with a topological phase transition at a nonzero $H_c$ to a vortex-free pinned phase.

PACS: 05.70.Jk, 64.60.Fr, 64.70.Pf, 75.50.Lk

There is considerable current interest in the effects of quenched disorder on ordered phases with continuous symmetry. Examples are vortex lattices in type-II superconductors [1], spin- and charge-density wave systems [2] subject to random pinning, as well as amorphous ferromagnets with random anisotropy [3] and liquid crystals in porous media [4]. The random pinning induces continuous, elastic distortions of the ordered state and it may also induce plastic deformation due to topological defects, such as dislocations, which do not represent continuous distortions of the ideal ordered state. The distortions induced by random pinning in the absence of topological defects can be treated within elasticity theory and have received some attention over the years [5–9]. Here we focus instead on the production of topological defects by the random pinning [10].

The simplest system in which to study these issues appears to be the ferromagnetic XY model with a random field. Here the long-range order in the pure system is ferromagnetism, the pinning is due to the random field, and the topological defects are vortices in the magnetization pattern. The hamiltonian we consider is

$$H = -\sum_{<i,j>} \mathbf{S_i} \cdot \mathbf{S_j} - \sum_i \mathbf{h_i} \cdot \mathbf{S_i} , \quad (1)$$

where the first sum is over all nearest-neighbor pairs of lattice sites, $\mathbf{S_i}$ is a unit-length, two-component (XY) spin at site $i$, and the static random fields $\mathbf{h_i}$ have rms magnitude $H$. The ground state of (1) evolves from ferromagnetic and vortex-free for $H = 0$ to a state with all spins aligned with the random fields, and thus a dense array of vortices, for large $H$. In the context of vortex lattices, the $H = 0$ limit corresponds to an unpinned Abrikosov lattice, while large $H$ corresponds to a vortex-glass ground state at strong pinning [1,11].

Let us first consider small random field $H$, following Imry and Ma [6]. Treating the random field as a perturbation, the long-range ordered state is stable at small $H$ only for dimension $d > 4$. For $d < 4$ the static elastic strains in the ground state induced by the random field are of order one at the pinning length $\xi_P \sim H^{-2/(4-d)}$. The behavior at longer scales is not accessible by simply perturbing in $H$. This perturbative treatment only considers continuous elastic distortions of the uniform state, not vortices, which are nonperturbative. Let us ask about the system's stability at small $H$ and length scale $L \leq \xi_P$ to static vortex/antivortex pairs in $d = 2$ or to vortex loops in $d = 3$. The vortices permit the system to align better with the random field on length scale $L$. This, naively, lowers the random field pinning energy by order $HmL^{d/2}$, where $m$ is the magnetization density at $H = 0$. However, the added energy of the elastic strains around the vortices is order $KL^{(d-2)}|\log(L)|$, where $K$ is the spin stiffness. Even at $L = \xi_P$, the elastic energy cost is larger than the pinning energy gain by a factor of $|\log(H)|$. Hence, the system appears to be stable against vortices for small $H$ at length scales less than or of order $\xi_P(H)$. Thus, we conclude that if the system has vortices at small $H$, the length scale at which the vortices appear, $\xi_V$, is larger than the pinning length, $\xi_P$. Other than this lower bound, how $\xi_V$ depends on $H$ and when and how the static vortices proliferate at scales $L > \xi_P$ is not known. This is what we investigate in this paper.

If vortices are forbidden, then the long-distance spin-spin correlation function is expected to decay as a power-law [9]. This behavior should also apply in the distance range between $\xi_P$ and $\xi_V$ where the system is strongly pinned but still largely vortex-free. Thus the true correlation length, beyond which the correlations decay exponentially, should be of order $\xi_V$. There are two possible behaviors for $\xi_V(H)$: (i) it may diverge only at $H = 0$, but with a stronger divergence than $\xi_P$, or (ii) it could diverge at a nonzero critical field, $H_c$. Our results below are more consistent with the former for $d = 2$ and the latter for $d = 3$. In the latter case there is an intermediate pinned phase for $0 < H < H_c$ that is vortex-free at the largest length scales, and therefore has topological long-range order. This novel ordered pinned phase, if it exists, has power-law long-distance spin-spin correlations distances and is separated from the disordered, plastic phase at higher $H$ by a topological phase transition at $H_c$ where large-scale static vortices first appear.

We have performed Monte Carlo simulations of the random-field XY model (1) on simple cubic ($d = 3$) and square ($d = 2$) lattices. In both cases we find that, as expected by the above argument, the spacing between vortices diverges more strongly with decreasing $H$ than the pinning length $\xi_P$ obtained from the Imry-Ma estimate. For $d = 3$, the data are consistent with a transition to a vortex-free phase at a nonzero critical field $H_c > 0$. For $d = 2$, on the other hand, the vortex density is better fit by a power-law in $H$, indicating that there is no intermediate vortex-free phase. For $d = 3$ we have tried to more precisely locate the proposed phase transition at $H_c$ using various forms of finite-size scaling involving moments of the magnetization distribution and the derivative of the magnetization with respect to $H$, but have been unsuccessful in obtaining convincing one-parameter scaling $\{L/\xi(H)\}$. This lack of good finite-size scaling may be because there are two distinct diverging length scales in this problem, namely $\xi_P$ and $\xi_V$, which complicates the finite-size scaling.

Here we report on simulations of large lattices (up to $10^6$ spins) in the field range that we could equilibrate. We simulated two copies (replicas) of each sample, one with a ferromagnetic initial condition and one with the spins initially aligned with the random field at each site. We only use late-time results where both replicas give the same time-independent averages. At the lowest fields studied, this required $10^5$ Monte Carlo steps per spin. We studied the model with Gaussian-distributed random fields $[\mathbf{h_i}] = \mathbf{0}$, $[\mathbf{h_i} \cdot \mathbf{h_j}] = H^2 \delta_{ij}$. We measured the time-averaged magnetization, $\mathbf{m_i} = <\mathbf{S_i}>$, at each site. The angle between the vectors $\mathbf{m_i}$ on each nearest-neighbor pair of sites was obtained, with the convention that it lies between $-\pi$ and $\pi$. For each elementary square plaquette, these angles were added to obtain the total rotation of the magnetization on moving around the plaquette. This sum is a multiple of $2\pi$; if it is nonzero, then there is a vortex in that plaquette in the equilibrium (static) magnetization pattern. We also measured the correlation function $g(r) = [\mathbf{m_i} \cdot \mathbf{m_j}]$, averaged on pairs of sites at distance $r$ along a lattice axis. We want to study the ordered-phase (low-temperature) behavior, but be at a high enough temperature that we can equilibrate in not too much computer time. For $d = 3$, where the critical point in the absence of the random field ($H = 0$) is at $T_c^{3D} \cong 2.2$ [12], we examined $T = 1.5$, which is enough below $T_c^{3D}$ to avoid the $H = 0$ critical regime. For $d = 2$, where $T_c^{2D} \cong 0.9$ [13], we studied $T = 0.7$.

The fraction, $f_V$, of elementary square plaquettes occupied by vortices is shown vs. $H$ in Fig. 1. The solid and dashed lines indicate what the slopes would be if the intervortex spacing was the pinning length, $\xi_P$, given by the Imry-Ma argument, so $f_V \sim \xi_P^{-2}$. The above argument says that $f_V$ should vanish more rapidly than this with decreasing $H$, and the data support this. For $d = 3$, the vortex density is not well approximated by a power of $H$ over any substantial field range.

In Fig. 2a we show the correlation function for $d = 3$, which is very well fit by a simple exponential: $g(r) \sim \exp(-r/\xi)$. The measured correlation length, $\xi$, also diverges faster than the Imry-Ma $\xi_P$ as $H$ is decreased. We find that the vortex density, $f_V$, equilibrates much earlier in the simulation than the correlation function, hence we have results that we trust for $f_V$ to lower field than for $g(r)$.

One naively expects that if there is a transition to a vortex-free phase for $d = 3$ at a nonzero $H_c$ (which depends on $T$), the vortex density $f_V$ vanishes and the correlation length $\xi$ diverges as a power of $(H - H_c)$. In Fig. 3 we show that our data for $T = 1.5$ are consistent with such a critical behavior with $H_c \cong 1.35$. The apparent exponents with $H_c = 1.35$, indicated by the solid lines in Fig. 3, are $\xi \sim (H - H_c)^{-\nu}$ with $\nu \cong 0.85$, and $f_V \sim (H - H_c)^\rho$ with $\rho \cong 1.4$. Note that the range of the scaling fits for all the quantities in Fig. 3 is less than one decade, so this apparent scaling should not be taken too seriously. But we can definitely say that the data for $f_V$ in this field range are very different from a power-law critical point with $H_c = 0$ and, with the data for $\xi$, consistent with a power-law critical point with $H_c$ near 1.3.

For $d = 2$ the Imry-Ma argument gives $\xi_P \sim 1/H$ at $T = 0$. At finite temperature, the magnetization at scale $L$, $m(L)$, is renormalized by thermal fluctuations in the critical phase below the Kosterlitz-Thouless temperature $T_{\mathrm{KT}}$. When the Imry-Ma argument is modified to take this into account one obtains $\xi_P \sim H^{-2/(2-\eta)}$. If we then take the largest value $\eta = 1/4$ at the Kosterlitz-Thouless transition temperature [14], we obtain the slope 16/7 indicated by the solid line near the $d = 2$ data in Fig. 1. As expected, the vortex density for $d = 2$ is found to vanish even faster than this with decreasing $H$. This again indicates that, as found in $d = 3$, the vortex spacing diverges more strongly than the pinning length given by the Imry-Ma argument. However, for $d = 2$ the low field behavior is quite consistent with a power-law in $H$, just with a larger exponent (roughly $f_V \sim H^3$), rather than a phase transition at nonzero $H_c$. The correlation function for $d = 2$, (Fig. 2b) and low field does not fit a simple exponential as found in $d = 3$, so it is not obvious how one should define the correlation length.

How should one think about the physics at length scales beyond $\xi_P$? One possibility is the following: Consider $d = 2$, for concreteness. First, forbid vortices and find the lowest energy state. This state is pinned with some particular nontrivial spin pattern. Now consider a patch of linear size $L$, (area $L^2$) with $L \gg \xi_P$. Introduce a vortex-antivortex pair in this patch and choose their locations and the spins' orientations to minimize the energy with this pair present. For $L < \xi_V$ this new energy is presumably typically higher than the lowest-energy vortex-free state, but for $L > \xi_V$ it is typically lower, so the true ground state has typical vortex spacing $\xi_V$.

What does the ground state with vortex separation $\xi_V$ look like? The *relative* spin orientation between it and the lowest-energy vortex-free state must rotate by $2\pi$ on encircling any vortex. But the system (for $\xi_V \gg \xi_P$) is strongly pinned, so it will typically cost a lot of energy to locally rotate the spins away from their local ground state. Thus we expect that the relative phase difference will typically be concentrated in line defects (like Sine-Gordon solitons or domain walls) that extend from vortex to antivortex. Thus to find the ground state one must optimize not only over the positions of the vortices but also of these line defects (these defects are present only in the *relative* phases and are presumably of width of order $\xi_P$). Since the line defect is not permitted in the vortex-free state, it can have negative energy relative to the lowest-energy vortex-free state once its position is optimized. This can then "pay for" the positive energy of the vortex cores. This picture provides an energy-balance mechanism for a phase transition driven by a proliferation of topological defects. The description of the physics at scale $L > \xi_P$ in this picture relies heavily on the existence of topological defects (vortices). It is not at all a renormalized elastic medium of the sort considered by Giamarchi and Le Doussal [9]. It is still unclear how to This scenario may also apply to random-*anisotropy* XY models [3,15].

We now comment on the connection between the $d = 3$ random field XY model and the Abrikosov vortex lattice with uncorrelated random pinning.

There are two aspects of the vortex lattice that are not captured by the XY model. The single continuous degree of freedom in the XY model, the spin orientation, plays the role of the vector displacement of the vortex lattice. One consequence of this simplification is that the topological charge of a vortex in the XY model is a scalar (how much the spin orientation winds upon encircling the vortex), while that of a dislocation in the vortex lattice is the two-component Burgers' vector. The other simplification is that the superconductor has the additional $U(1)$ symmetry of the complex scalar Ginzburg-Landau parameter $\psi$ under rotations in the complex plane; this symmetry is absent in the random-field XY model. This $U(1)$ symmetry can be spontaneously broken in the absence of any vortex-lattice order in the superconductor, yielding the vortex glass phase [1,11]. The random-field XY model has no analogous ordered phase. Another consequence of the U(1) sym-metry is that in the vortex lattice an interstitial (or vacancy), being a change in the vortex number, is itself a topological defect; the random-field XY model has no analogous defect.

The vortex-free pinned phase at intermediate disorder $H$ that we discuss in this paper for the random field XY model corresponds to a pinned, superconducting, dislocation-free Abrikosov vortex lattice phase in a type-II superconductor [11]. In the latter system, the structure factor, $S(\mathbf{q})$, would have power law singularities at the basic reciprocal lattice vectors, $\mathbf{Q}$, of the form $S(\mathbf{q}) \sim |\mathbf{q} - \mathbf{Q}|^{-(2-\eta)}$ [9]. By increasing the disorder $H$ in the random-field XY model, we drive the system into the fully disordered phase. In that case, the disordered phase is not thermodynamically distinct from the high-temperature paramagnetic phase. For the superconductor, on the other hand, there are at least two noncrystalline phases: the superconducting vortex glass phase at low $T$ and the resistive vortex liquid phase at higher $T$. The transition directly from the vortex lattice phase to the vortex liquid involves loss of both crystalline and superconducting order, so is not properly modelled by the random-field XY model. In the superconductor, this melting transition is first-order; the XY model shows no such transition. However, for the transition from the pinned vortex lattice to the vortex glass, both phases have an off-diagonal long-range (superconducting) order, so the superconducting order could be merely a bystander, and the random-field XY model, which ignores this order, might capture the essential physics of this transition.

There is some evidence that there may be two distinct superconducting phases in clean crystals of YBCO and BSCCO. At low applied magnetic fields, transport measurements show a first-order melting transition of the vortex lattice [16,17]. At higher fields, the effective random pinning appears stronger [18], and the superconducting transition is continuous, as expected for the melting of a vortex glass. For BSCCO, a neutron scattering study saw the lattice Bragg peaks in the low-field regime but not in the high-field regime [19]. This is all consistent with these samples having the pinned vortex lattice phase in the low field regime and the vortex glass phase in the high field regime. Some signs of a transition between these two superconducting phases have recently been seen in the nonlinear transport properties near the critical current in YBCO [20]. This transition, we argue, is characterized by the proliferation of dislocations in the lattice, reducing it to the amorphous vortex glass. It is this transition that is modelled by the random-field XY model we have studied here.

In conclusion, we have performed for the first time extensive Monte Carlo simulations of the $d = 2$ and $d = 3$ random field XY model. In both dimensions, the spacing between static vortices grows faster than the pinning length obtained from the Imry-Ma argument as the random field amplitude goes to zero. In $d = 3$, our results are consistent with a phase transition at nonzero critical random field into a novel topologically ordered (vortex-free) phase with power-law decay of the spin-spin correlation function.

We thank Sue Coppersmith for discussions. The work at TRIUMF was supported by the NSERC of Canada.

FIGURES

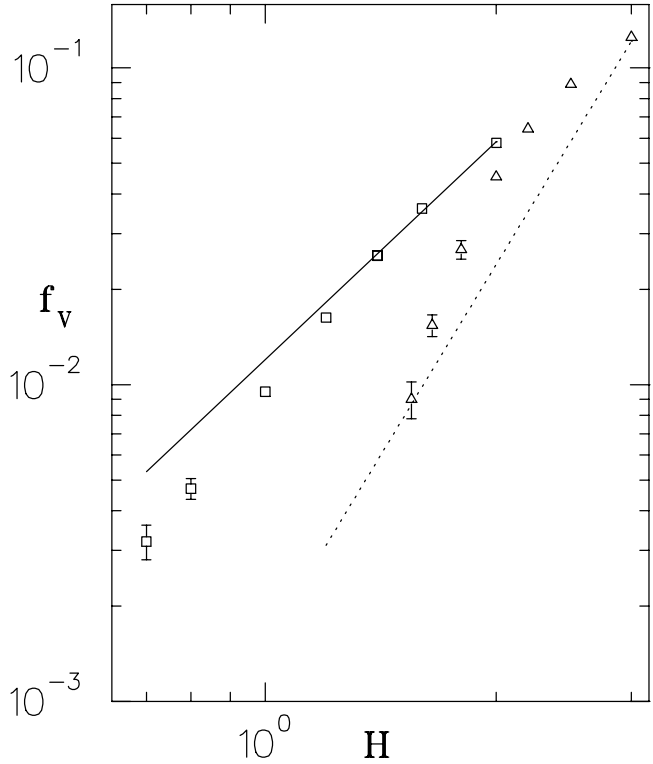

FIG. 1. The fraction of plaquettes occupied by vortices vs. the rms random field strength. Triangles are $d=3$, $T=1.5$; squares are $d=2$, $T=0.7$. The slopes of the lines on this log-log plot are 4 and 16/7. These data are from simulations of single large samples ($10^5 - 10^6$ spins), so the sample-to-sample statistical errors have not been properly estimated. From the cases where we did simulate more than one sample, we expect that the errors are roughly at the 5% level for the lowest fields and smaller at higher fields.

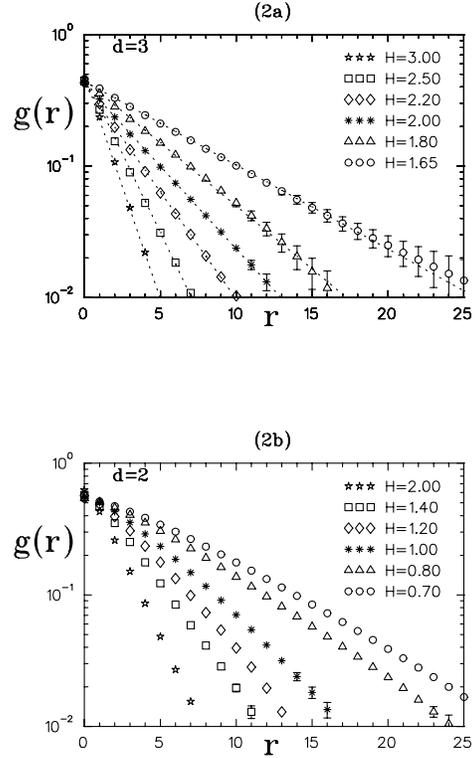

FIG. 2. The magnetization correlation function versus distance for (a) $d=3$, $T=1.5$, and (b) $d=2$, $T=0.7$, for the indicated random field strengths. The dotted lines in (a) are fits to simple exponentials. The error bars, where shown, indicate variations between 2-3 samples. Sample-to-sample statistical errors, although not properly measured, appear to be larger, particularly at low field and large $r$.

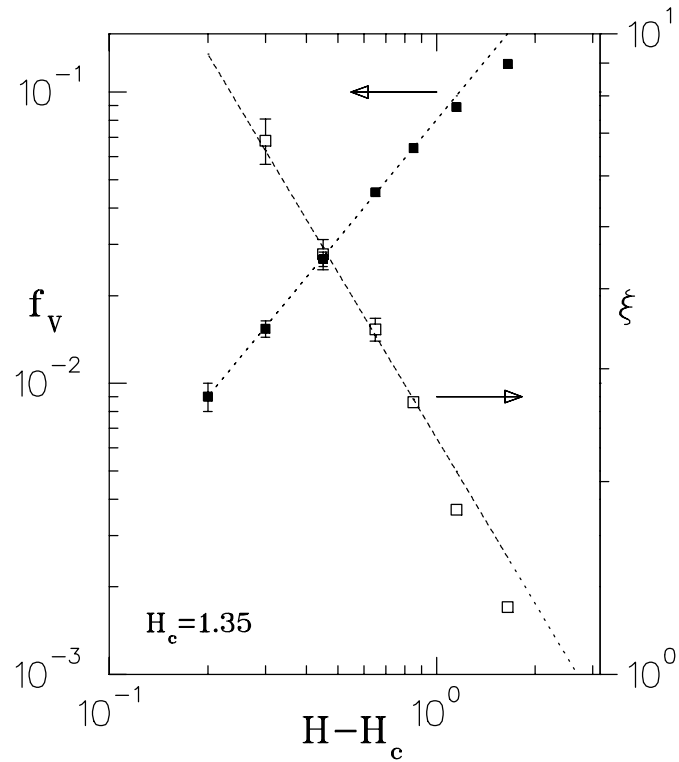

FIG. 3. For $d = 3$, $T = 1.5$ and $H_c = 1.35$ the vortex density, $f_V$, (solid symbols) and correlation length, $\xi$, (open symbols) vs. $(H - H_c)$ on a log-log plot. $\xi$ is obtained from the simple exponential fits in Fig. 2a.